\documentclass[prl,aps,twocolumn,amssymb,superscriptaddress,showpacs,floatfix]{revtex4-1}
\usepackage{amssymb}

\usepackage[pdftex]{graphicx}

\newcommand{\magnetit}{Fe$_3$O$_4$ }

\begin{document}

\title{Anharmonicity due to Electron-Phonon Coupling in Magnetite}

\author{Moritz Hoesch}
\affiliation{Diamond Light Source, Harwell Campus, Didcot OX11 0DE, Oxfordshire, England}

\author{Przemys\l{}aw Piekarz}
\affiliation{Institute of Nuclear  Physics, Polish Academy of Sciences,
             Radzikowskiego 152, PL-31342 Krak\'ow, Poland }

\author{Alexey Bosak}
\affiliation{European Synchrotron Radiation Facility, 6 rue Jules Horowitz,
             F-38043 Grenoble Cedex, France}

\author{Mathieu Le Tacon}
\affiliation{Max-Planck-Institut f\"ur Festk\"orperforschung,
             Heisenbergstrasse 1, D-70569 Stuttgart, Germany}

\author{Michael~Krisch}
\affiliation{European Synchrotron Radiation Facility, 6 rue Jules Horowitz,
             F-38043 Grenoble Cedex, France}

\author{Andrzej Koz\l{}owski}
\affiliation{Faculty of Physics and Applied Computer Science,
             AGH-University of Science and Technology, \\
             Aleja Mickiewicza 30, PL-30059 Krak\'ow, Poland}

\author{Andrzej M. Ole\'{s}}
\affiliation{Max-Planck-Institut f\"ur Festk\"orperforschung,
             Heisenbergstrasse 1, D-70569 Stuttgart, Germany}
\affiliation{Marian Smoluchowski Institute of Physics, Jagellonian University,
             Reymonta 4, PL-30059 Krak\'ow, Poland}

\author{Krzysztof Parlinski}
\affiliation{Institute of Nuclear Physics, Polish Academy of Sciences,
             Radzikowskiego 152, PL-31342 Krak\'ow, Poland }

\date{\today}

\begin{abstract}
We present the results of inelastic x-ray scattering for magnetite
and analyze the energies and widths of the phonon modes with
different symmetries in a broad range of temperature $125<T<293$ K. The
phonon modes with $X_4$ and $\Delta_5$ symmetries broaden in a nonlinear
way with decreasing $T$ when the Verwey transition is
approached. It is found that the maxima of phonon widths occur away from
high-symmetry points which suggests the incommensurate character of
critical fluctuations. Strong phonon anharmonicity induced by
electron-phonon coupling is discovered by a combination of these
experimental results with {\em ab initio}
calculations which take into account local Coulomb interactions at Fe
ions. It (i) explains observed anomalous phonon broadening, and (ii)
demonstrates that the Verwey transition is a cooperative phenomenon
which involves a wide spectrum of phonons coupled to the electron
charge fluctuations condensing in the low-symmetry phase.
\end{abstract}

\pacs{63.20.dd, 63.20.dk, 71.30.+h, 75.25.Dk}

\maketitle

Discovered in ancient Greece, magnetite played a crucial role in the
history of magnetism. Apart from its magnetic properties, which found
numerous technological applications during the last century, magnetite
exhibits extraordinary behavior connected with the Verwey transition
\cite{verwey}. The most prominent feature observed upon lowering the
temperature is the discontinuous reduction of the electric conductivity
by two orders of magnitude at $T_V=122$ K. Extensive experimental and
theoretical studies of magnetite during last decades revealed very
complex interrelations between the electronic and structural degrees of
freedom, which both participate in the Verwey phase transition
\cite{imada,review}. In spite of this progress, the transition is not
yet fully understood and remains a very intriguing phenomenon
in condensed matter physics.

At the Verwey transition, the crystal structure changes from the
high-temperature cubic $Fd{\bar 3}m$ to the low-temperature monoclinic
$Cc$ symmetry \cite{iizumi}. It was first believed that octahedral $B$
sites, with  mixed-valence Fe$^{2.5+}$ ions turn to distinct Fe$^{2+}$
and Fe$^{3+}$ ions below $T_V$. Detailed diffraction studies of the
low-temperature phase revealed a much more complex
charge-order pattern \cite{attfield}, with a wide distribution of
valences at the octahedral sites \cite{blasco}, and an orbitally ordered
pattern distributed over three Fe sites (trimerons) \cite{senn}. These
observations are consistent with {\it ab initio} calculations
\cite{leonov,jeng} and resonant x-ray scattering studies
\cite{nazarenko,schlappa,tanaka,Sub12}, that additionally suggest the
charge-orbital (CO) order appearing a few Kelvins above $T_V$
\cite{lorenzo} and possibly existing in the dynamic form even at higher
temperatures \cite{pontius}. Since the existence of CO order above the
Verwey transition has been questioned by other studies \cite{Gar09} and
its role in this transition is under debate at present \cite{Sub12},
experiments performed above $T_V$ might resolve this controversy.

The key element, which stabilizes the CO order and drives the crystal
symmetry change is the coupling between electronic and vibrational
degrees of freedom. Such cooperative nature of the Verwey transition
is supported by numerous experimental observations: oxygen isotope
effect \cite{terukov}, critical softening of the $c_{44}$ elastic
constant \cite{schwenk}, critical diffuse scattering
\cite{fujii,shapiro}, phonon anomalies measured by the Brillouin
\cite{seikh}, Raman \cite{gasparov}, and nuclear inelastic scattering
\cite{handke,kolodziej}.
The instability of the electronic structure is intimately connected with
a lattice deformation, with certain similarity to the Peierls model
\cite{Shc09}.
In the simplest model, one can explain the doubling of the unit cell
along the c direction by the condensation of the acoustic $\Delta_5$
mode at ${\vec q}=(0,0,\frac{1}{2})$ \cite{ihle}, whereas the monoclinic
distortion requires at least two primary order parameters with
$\Delta_5$ and $X_3$ symmetries \cite{prl}.

In this Letter, we report a lattice dynamics study of magnetite over a
wide range of temperatures above $T_V$ and address the question whether
or not the electron-phonon coupling in the presence of CO fluctuations
is the principal cause of the Verwey transition. Our study reveals
increasing anomalous broadening of the lowest transverse acoustic (TA)
modes along the [001] ($\Delta$) and transverse optic (TO) modes along
the [110] ($\Sigma$) directions as temperature decreases towards $T_V$.
We show by density functional theory (DFT) calculations that the
anharmonic phonon behavior discovered here:
(i) is a direct consequence of the electron-phonon coupling, and
(ii) occurs only in presence of strong electronic correlations which is
a novel aspect of the Verwey transition.

\begin{figure}[t!]
\centerline{\includegraphics[width = 6.9cm]{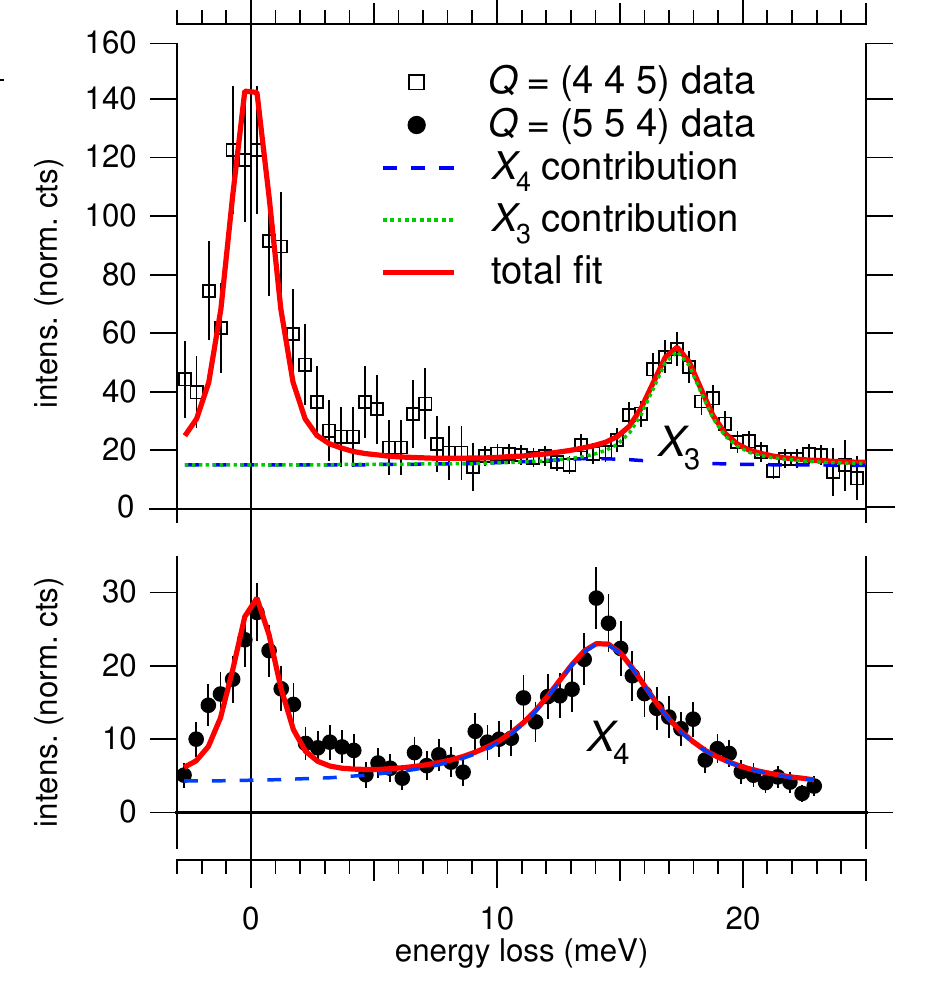}}
\vspace{-2mm}
\caption{Inelastic x-ray scattering spectra of \magnetit at $T= 150$ K.
At the $X$-point (554) scattering from the $X_4$ phonon and at (445)
from the $X_3$ phonon with a small contribution from $X_4$ are dominant.
Energy resolution is 1.6 meV.}
\label{Fig1}
\end{figure}

Momentum resolved vibrational spectra were obtained by
inelastic x-ray scattering (IXS)
at beamline ID28 at the European Synchrotron Radiation Facility (ESRF)
\cite{krisch07}. Two settings with energy resolution of $\delta E = 1.6$
meV and 3.2 meV were used. The sample was a single crystal of magnetite,
skull melter grown at Purdue University and annealed for stoichiometry
\cite{harrison}. The sample temperature was measured using a Si diode
sensor on the copper cold finger of the closed cycle cryostat.

The momentum points $Q$ to be measured were chosen from the calculations
presented below. From the calculated phonon Eigenvectors the IXS
intensity was predicted and favorable $X$-points $Q$ were selected from
the highest contrast against other modes and high total intensity. The
most suitable $X_4$ mode was found at $Q = (554)$. The optimal $X_3$
mode was found at $Q=(445)$ with a slight contribution from the $X_4$
mode in the spectra. The lowest optical $\Delta_5$ mode was measured
along the line from $Q=(333)$ to $Q=(334)$.

\begin{figure}[t!]
\centerline{\includegraphics[width = 8.2cm]{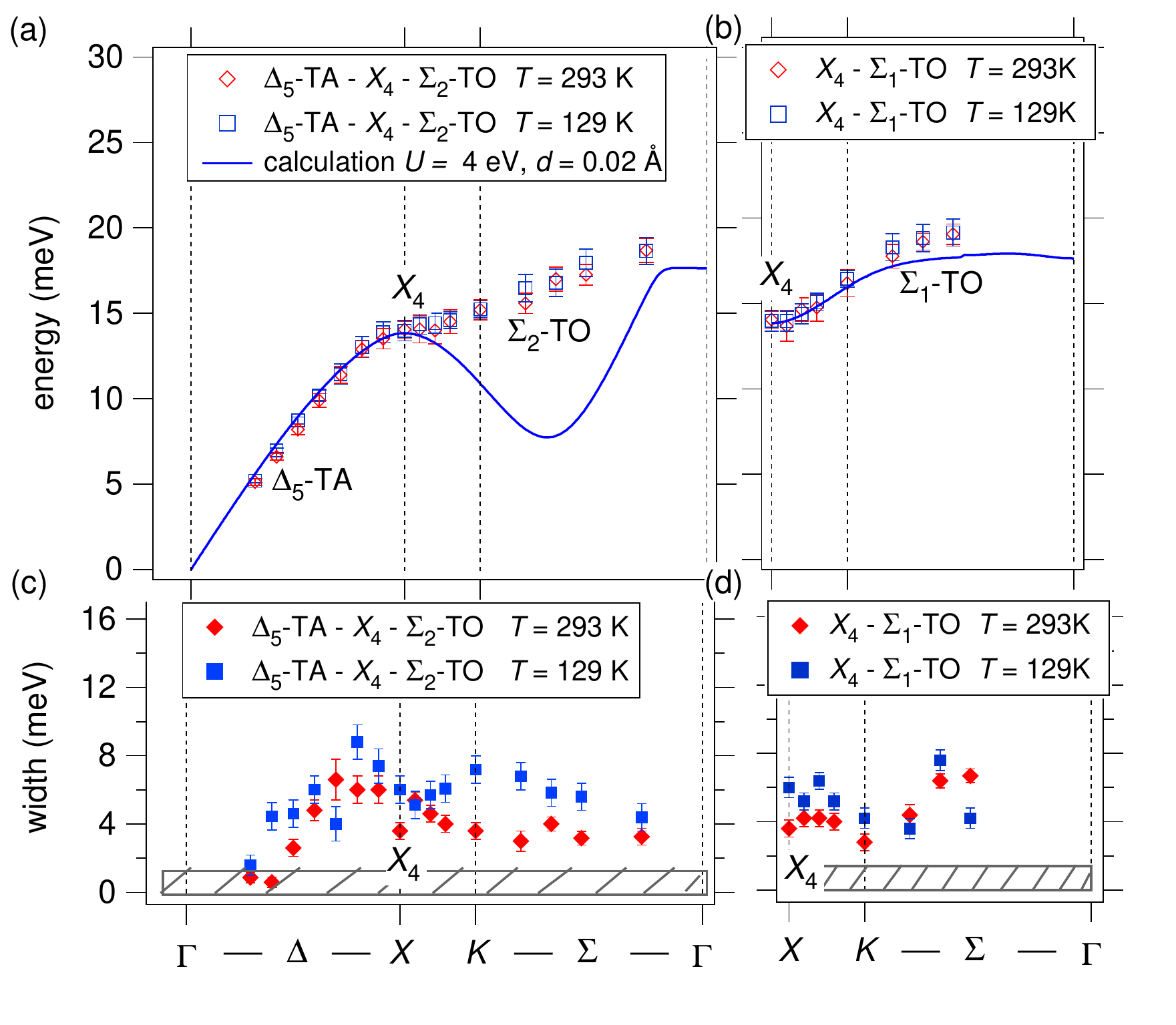}}
\vspace{-3mm}
\caption{(a,b) Phonon energies determined by peak fitting as a
function of reduced wave vector for branches connected to $X_4$,
measured in the vicinity of (554), for two temperatures $T=293$~K and
$T=129$~K. Calculated phonon dispersions are plotted with solid lines.
(c,d) Phonon widths of branches connected to $X_4$ as a
function of reduced wave vector.}
\label{Fig2}
\end{figure}

Raw data with $\delta E=1.6$ meV at the Brillouin zone boundary point $X$
are shown in Fig.~\ref{Fig1}. The lowest energy $X_4$ phonon displays a
significant broadening, while the lowest $X_3$ phonon is only slightly
broader than the experimental resolution. As the temperature is
lowered towards $T_V$ the diffuse scattering intensity at zero energy
loss  strongly increases. The phonon energies on the other hand show no
readily visible temperature dependence.

To extract precise phonon energies and peak widths the spectra were
numerically fitted by a superposition of peaks derived from the measured
resolution function. For visibly enlarged phonon peaks these were
convoluted with a Lorentzian function. For the quasi-elastic intensity
no convolution was applied.
The spectra at (554) of Fig.~\ref{Fig1} were thus analyzed first, and
the determined phonon energy and width were kept fixed
for the fit of the data at (445). The free parameters were the
intensities and one phonon energy and width per spectrum. We found that
the broadening had to be applied to the $X_4$ mode and all connected
branches, while the $X_3$ mode and connected branches remained close to
resolution limited.

\begin{figure}[t!]
\centerline{\includegraphics[width = .38\textwidth]{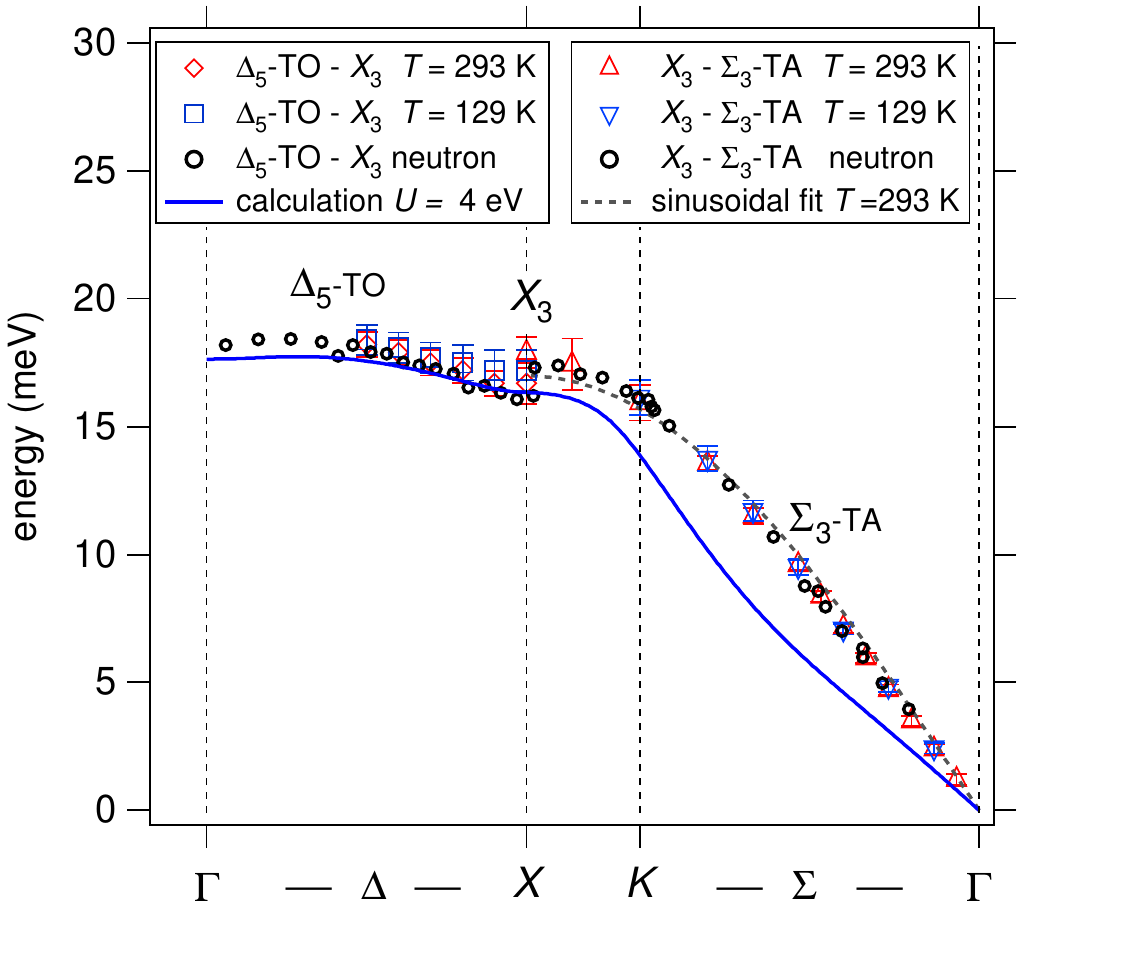}}
\vspace{-3mm}
\caption{Phonon energies as a function of reduced wave vector
for branches connected to $X_3$,
measured in the vicinity of (445) and (334).
Data at two temperatures $T=293$~K and $T=129$~K are shown.
Calculated phonon dispersions are plotted with solid lines. Data from
neutron scattering at room temperature~\cite{samuelsen} and a sinusodal
fit to represent a typical phonon dispersion are also shown.
}
\label{Fig3}
\end{figure}

The thus determined phonon dispersions from data at $\delta E = 3.2$ meV
are shown in Figs.~\ref{Fig2}(a,b) for branches connected to $X_4$ and
in Fig.~\ref{Fig3} for $X_3$. Fig.~\ref{Fig3} also includes room
temperature inelastic neutron scattering data \cite{samuelsen} and the
acoustic $\Sigma_3$-TA mode is fit by a model sinusoidal dispersion,
which reveals a downward bending of the dispersion by less than 1 meV
about half way between $K$ and $\Gamma$.
The dispersions match within the error bars both at room temperature and
$T=129$ K, with only a slight hardening at low temperature visible for
the $X_4$ mode, in agreement with the measured Fe density of states
\cite{handke,kolodziej} (below $T_V$ the multi-twinned nature of the
sample prevents the determination of the phonon dispersion).
In the following the $\Delta_5$-TA, $\Sigma_1$-TO and
$\Sigma_2$-TO will be denoted as the branches connected to $X_4$,
and $\Delta_5$-TO and $\Sigma_3$-TA as connected to $X_3$.
In contrast, the phonon widths in Figs.~\ref{Fig2}(c) and \ref{Fig2}(d)
have a much stronger temperature dependence. The width increases on
lowering the temperature which contradicts the usual situation, where
anharmonicity increases with increasing
temperature due to phonon-phonon interactions.

\begin{figure}[t!]
\centerline{\includegraphics[width = 6.8cm]{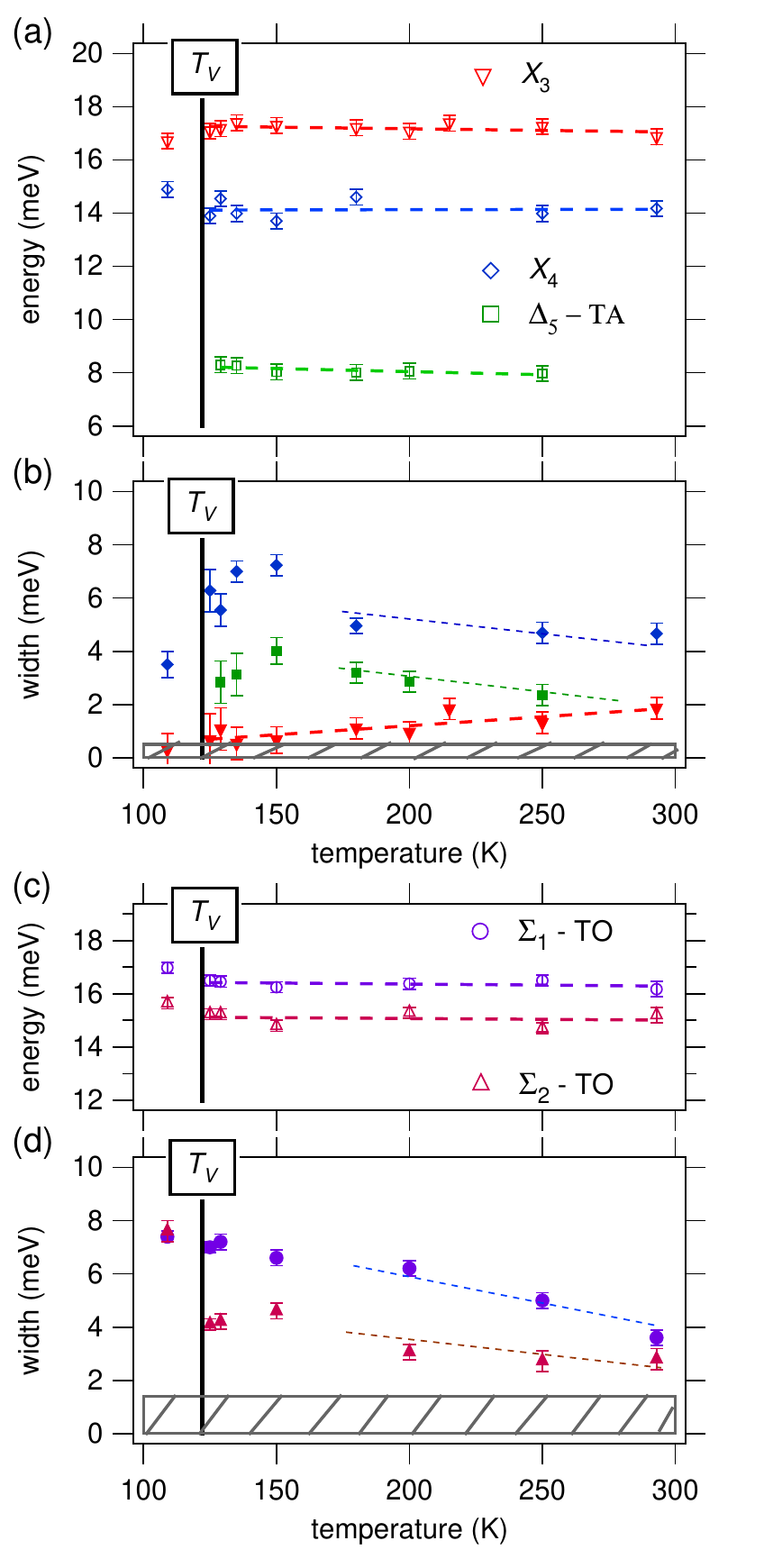}}
\vspace{-2mm}
\caption{(a,c)
Energies of phonon modes at the $K$-point as a function of temperature:
(a) $\Delta_5$, $X_4$ and $X_3$, and
(c) $\Sigma_1$ and $\Sigma_{2}$;
Widths of these phonon modes
are displayed in (b) and (d), respectively.
Straight dashed lines in (b) and (d) were fitted to points at high
temperature. Widths within
the hatched area are not reliably extracted from the data.}
\label{Fig4}
\end{figure}

The full temperature dependence of selected modes, measured with
$\delta E = 1.6$ meV is shown in Fig.~\ref{Fig4}. Remarkably, the modes
connected to $X_4$ show an increasing width on lowering the temperature
towards $T_V$. The $X_4$ and $\Delta_5$-TA phonons show already
anomalous widths at room temperature moderately increasing with
decreasing temperature. After reaching maximum values around 150 K,
their widths are reduced again in a temperature range of about 20 K
above $T_V$. The $\Sigma_1$ and $\Sigma_2$ modes, measured at the $K$
point, show even stronger and linear increase of widths over the entire
range of temperatures.

The DFT calculations were performed using the {\sc vasp} program
\cite{vasp} within the generalized gradient approximation (GGA) and
the full-potential projector-augmented wave method \cite{paw}. Phonon
energies for the cubic $Fd\bar{3}m$ symmetry were calculated in the
$1\times 1\times 1$ supercell with 56 atoms using the direct method
\cite{direct}, implemented in the {\sc phonon} program \cite{phonon}.
On the basis of the Hellmann-Feynmann theorem, the atomic forces
were obtained by displacing atoms from their equilibrium positions.
For the cubic symmetry only three independent displacements of Fe($A$),
Fe($B$), and O atoms are sufficient to derive all force-constants
matrix elements and the dynamical matrix. In order to study anharmonic
effects, we have calculated the Hellmann-Feynmann forces
and phonon energies for two independent sets of displacements with
the amplitudes $u=0.02$ and $0.04$ \AA. The effect of local electron
interactions on phonon energies was investigated by performing
calculations within the GGA+$U$ approach, where the on-site Coulomb
interaction and Hund's exchange parameters, $U=4.0$ eV and $J=0.8$ eV,
are the same as in the previous studies \cite{prl}.

Considering the $\Delta$ and $\Sigma$ directions,
we found a very good agreement for both the $\Delta_5$ modes between the
experimental points and the calculated phonon dispersions for $u=0.02$
\AA\, and only a small shift of the $\Sigma_1$ mode to lower energies
(Fig. \ref{Fig2}).
The largest discrepancy is found for the $\Sigma_2$ mode,
which shows anomalous softening induced by the local electron
interactions \cite{prl}, directly connected with the anharmonic
behavior discussed below. The $\Sigma_3$ mode exibits anomalous
downward bending, much larger than those observed in the experiment.

\begin{figure}[t!]
\centerline{\includegraphics[width = 7.6cm]{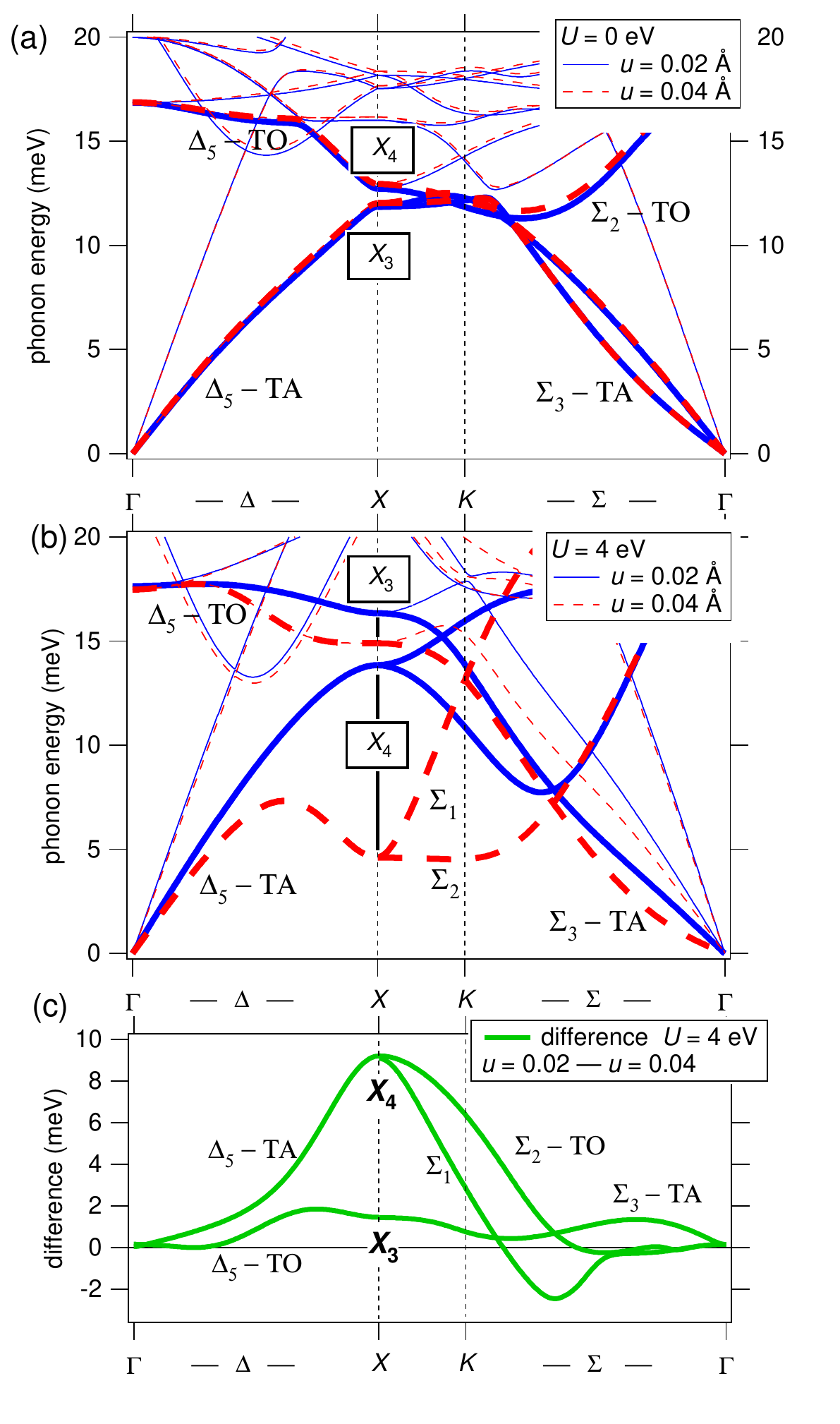}}
\vspace{-3mm}
\caption{Calculated phonon dispersions of \magnetit
obtained from GGA+$U$ combined with the {\sc phonon} software \cite{phonon},
for: (a) $U=0$, and (b) $U=4$~eV,
shown by solid (dashed) lines for a displacement of $u = 0.02$~\AA{}
($0.04$~\AA).
Selected branches connected to the lowest $X_3$ and $X_4$ optical modes
are highlighted. The difference between the phonon dispersions obtained
at both values of $u$ for $U = 4$ eV is shown in (c).
}
\label{Fig5}
\end{figure}

Coulomb interactions strongly influence phonon dispersions.
For $U=0$ [Fig. \ref{Fig5}(a)] the phonon energies obtained
for $u=0.02$ and $0.04$ \AA{} are very close to each other
and differ by less than 0.5 meV (not shown),
which means slight hardening of phonons for increasing displacement
amplitudes. It demonstrates that the system is harmonic and phonon
frequencies depend only very weakly on vibrational amplitudes.
The situation changes in a dramatic way in presence of local electron
interactions [see Figs. \ref{Fig5}(b,c)] --- for $U=4$ eV the phonon
energies depend strongly on $u$ and substantially soften for $u=0.04$
\AA. The largest changes are observed for the branches connected to
$X_4$ with a broad maximum in the energy differences at the $X$ point.
A weaker but still pronounced effect is visible also for branches
connected to $X_3$.

We emphasize that the same modes which have anomalous temperature
dependence of phonon widths exhibit strong anharmonic behavior in the
theory when $U=4$ eV. Moreover, the phonon widths for the $\Delta_5$-TA
and $\Sigma_2$-TO modes (Fig. \ref{Fig2}) show similar trends as the
plot of energy differences, see Fig. \ref{Fig5}(c). In both cases
the largest values are located in the vicinity of the $X$ point. Along
the $\Delta$ direction, one finds a pronounced increase of values in
both figures around ${\vec q}=(0,0,\frac12)$.
The phonon widths at room temperature have
a maximum around ${\vec q}=(0,0,0.7)$ and decrease systematically when
${\vec q}$ increases towards the $X$ point,
and further along the $\Sigma$ direction. At low temperature a two-peak
structure for the ${\vec q}$-dependent width is apparent: one peak at
${\vec q}\simeq (0,0,0.8)$ and the other one at the $K$ point.

The observed temperature dependence of the phonon widths
and the exact positions of their maxima cannot be properly
described within the present calculations, see Fig. \ref{Fig5}(c).
The shift of peaks away from the $X$ point indicates the existence
of incommensurate correlations, which are missing in the supercell
calculations due to the geometrical constraints.

In standard theories, large phonon widths indicate either an anharmonic
behavior or strong electron-phonon coupling \cite{abrikosov}.
Our {\it ab initio} study demonstrates that these two effects are
intimately connected in magnetite and both result from local electron
interactions. These interactions induce polarization of the
minority-spin partly occupied $t_{2g}$ orbitals.
In such a correlated state, the coupling between electrons
and lattice enhances and stabilizes the CO order,
effectively reducing the total energy \cite{prl,pinto}.
This in turn modifies the interatomic interactions and generates the
anharmonic potential. Indeed, an anharmonic double-well potential
has been obtained for the $\Delta_5$ and $X_3$ modes
in the presence of strong electron correlations
\cite{prl}. This mechanism leads to the Verwey transition with the
static structural distortion and frozen CO order below $T_V$.
In contrast, above $T_V$ the CO degrees of freedom are dynamic and
couple strongly
to phonons reducing their lifetimes. Therefore we conclude that the
anharmonicity in magnetite is induced by the coupling between phonons
and CO fluctuations which leads to phonon broadening \cite{noteline}.

We remark that a strong electron-phonon coupling is consistent with the
polaronic short-range order observed above $T_V$
\cite{ihle2,park2,schrupp}.
The anomalous phonon broadening correlates also very well with the
critical diffuse scattering existing over a broad region in reciprocal
space and over a broad temperature range \cite{shapiro}.
In both cases, the strongest response is observed at incommensurate
wavelengths in the reciprocal space.
Charge fluctuations observed in a wider temperature range may be
responsible for distinct properties of the Verwey transition in
non-stoichiometric or doped magnetite \cite{aragon,kozlowski,kolodziej}.

In summary, we have presented the IXS measurements of the lowest phonon
modes along the $\Delta$ and $\Sigma$ directions in magnetite
above the Verwey transition --- they reveal an anomalous behavior of
phonon widths that {\it increase\/} with lowering temperature.
Their maxima are found away from the high-symmetry points due to the
incommensurate character of critical fluctuations. Within the
{\it ab initio} study we have found that the same phonon modes show
strong softening for larger amplitudes in the presence of local
electron interactions in the Fe($3d$) states. This phonon anharmonicity
is induced by the charge-orbital fluctuations above $T_V$
--- it demonstrates that the Verwey transition is triggered
by a cooperative electron-phonon mechanism at large $U$ \cite{prl}.

{\it Acknowledgments.---}
We wish to thank K. Refson, L.F.~Feiner, P.T. Jochym, and
J. \L{}a\.zewski for insightful discussions and D. Gambetti and
P. Dideron for technical assistance. This work was performed at the
European Synchrotron Radiation Facility.
P.P. and A.M.O. acknowledge support by the
Polish National Science Center (NCN) under Projects
No. 2011/01/M/ST3/00738 and No. 2012/04/A/ST3/00331.


\begin{thebibliography}{99}
\bibitem{noteline} Interestingly, anharmonic effects/phonon line broadening
                are visible for the $\Delta_5$ phonon mode and not for the
                $X_3$ mode, which strongly couples to the $t_{2g}$ states
                at the Fermi energy and opens the insulating gap \cite{prl}.
\end{thebibliography}
\end{document}